\documentstyle[12pt,epsf]{article}

\begin{document}        
\pagestyle{empty}
\renewcommand{\thefootnote}{\fnsymbol{footnote}}

\begin{flushright}
{\small
SLAC--PUB--8644\\
Oct 2000\\}
\end{flushright}
 
\vspace{.8cm}


\begin{center}
{\bf\large   
Measurements of Parity-Violation Parameters at SLD\footnote{Work supported 
by Department of Energy contract  DE--AC03--76SF00515 (SLAC).}}

\vspace{1cm}
{\bf
Masako Iwasaki
} \\
\vskip 0.2cm
{\it
Department of Physics, University of Oregon, Eugene,
OR 97403
} \\
\vskip 0.6cm
{\bf
Representing the SLD Collaboration$^{**}$
} \\
\vskip 0.2cm
{\it
Stanford Linear Accelerator Center, Stanford University,
Stanford, CA  94309
}
\medskip
\end{center}
 
\vfill

\begin{center}
{\bf\large   
Abstract }
\end{center}
We present direct measurements of the parity-violation parameters 
$A_b$, $A_c$, and $A_s$ at the $Z^0$ resonance 
with the SLD detector.
The measurements are based on approximately 530k hadronic $Z^0$ events
collected in 1993-98.
Obtained results are
$A_b = 0.914\pm0.024$ (SLD combined: preliminary), 
$A_c = 0.635\pm0.027$ (SLD combined: preliminary), 
and $A_s = 0.895\pm0.066(stat.)\pm 0.062(sys.)$.
\vfill

\begin{center} 
{\it Presented at the XXXth International Conference 
On High-Energy Physics (ICHEP 2000), 27 Jul - 2 Aug 2000, 
Osaka, Japan}
\end{center}

\newpage
\pagestyle{plain}

\section{Introduction}               
In the Standard Model, the $Z^0$ coupling to fermions has 
both vector ($v_f$) and axial-vector ($a_f$) components. 
Measurements of fermion asymmetries at the $Z^0$ resonance probe
a combination of these components given by 
$ A_f = 2v_f a_f/(v_f^2 + a_f^2)$.
The parameter $A_f$ expresses the extent of parity violation 
at the $Zf\bar{f}$ vertex
and its measurement provides a sensitive test of the Standard Model.

At the Stanford Linear Collider (SLC), 
the ability to manipulate the longitudinal polarization 
of the electron beam allows the isolation
of $A_f$ 
through formation of the left-right forward-backward
asymmetry:
$$
\tilde{A}_{FB}^f(z)=
\frac
{[\sigma_L^f(z)-\sigma_L^f(-z)]-[\sigma_R^f(z)-\sigma_R^f(-z)]}
{[\sigma_L^f(z)+\sigma_L^f(-z)]+[\sigma_R^f(z)+\sigma_R^f(-z)]}
=|P_e|A_f\frac{2z}{1+z^2},
$$
where $P_e$ is the longitudinal polarization of the electron beam, 
and $z = \cos\theta$ is the 
direction of the outgoing fermion relative to the incident electron.

The measurements described here are based on 530k $Z^0$-decay events
taken in 1993-98 with the SLC Large Detector (SLD)\cite{sld}.
The average electron polarization is $|P_e| = 73 \pm 0.5\%$\cite{SLDALR}.
Polarized electron beams, a small and stable SLC interaction region,
the high resolution CCD vertex detector\cite{vxd3}, and 
the excellent particle identification 
with \v{C}erenkov Ring imaging Detector (CRID)\cite{crid}
provide precision electroweak measurements.

\section{$A_b$ measurements}
In order to tag the $b$-quark, 
topologically reconstructed mass of the secondary vertex\cite{masstag}
is used.
The secondary vertex is  reconstructed with charged tracks,
and its invariant mass is calculated.
To account for neutral particles and missing tracks,
the vertex mass is corrected: 
we calculate the $P_T$-corrected mass $M_{P_T}$
by estimating a missing $P_T$ from the acolinearity between
the momentum sum of the vertex and the direction of the 
vertex flight path.
Applying the cut of $M_{P_T}>2\ GeV/c^2$, 
we identify the $b$-quark with 98\% purity and 50\% efficiency.

To determine the $b$-quark charge, we uses 4 different methods:
1) vertex charge, 2) jet charge, 3) cascade kaon
and 4) lepton. 

The vertex-charge analysis uses the track charge sum of the 
secondary vertex to identify the charge of the 
primary quark\cite{SLDAbVTX}.
We introduce the
Neural Network technique to reject background and 
to associate the tracks to the secondary vertices. 
It improves the $b$-tagging efficiency to 57\%.
In this analysis, we reconstruct the tracks which has hits 
in the vertex detector only.
By adding such tracks, we enhance the charge separation performance
to 83\%. 
The $b$-tagging purity and correct charge probability
are estimated using opposite hemisphere information.

In the jet-charge analysis, we use the net momentum-weighted 
jet-charge\cite{SLDAbJet}.
The track charge sum and difference between the two
hemispheres are used to extract the analyzing power from data, thereby
reducing MC dependencies and lowering systematic effects.

In the kaon analysis, we use the charged kaon in 
the decay ${\bar{B}} \rightarrow D
\rightarrow K^-$, to determine the $b$ charge\cite{SLDAbK}.
CRID is used to identify $K^\pm$ with high-impact parameter
tracks. 
The charges of the
kaon candidates are summed in each hemisphere and the difference
between the two hemisphere charges is used to determine the polarity
of the thrust axis for the $b$-quark direction.

Electrons and muons are used to identify the 
charge and direction of the primary $b$ quark\cite{SLDAbAclept}.
Geometrical information is used to separate cascade and prompt
leptons. In the electron analysis, we also use the Neural Network 
for source classification.

\begin{figure}
\begin{center}
\epsfysize 14cm
\epsfbox{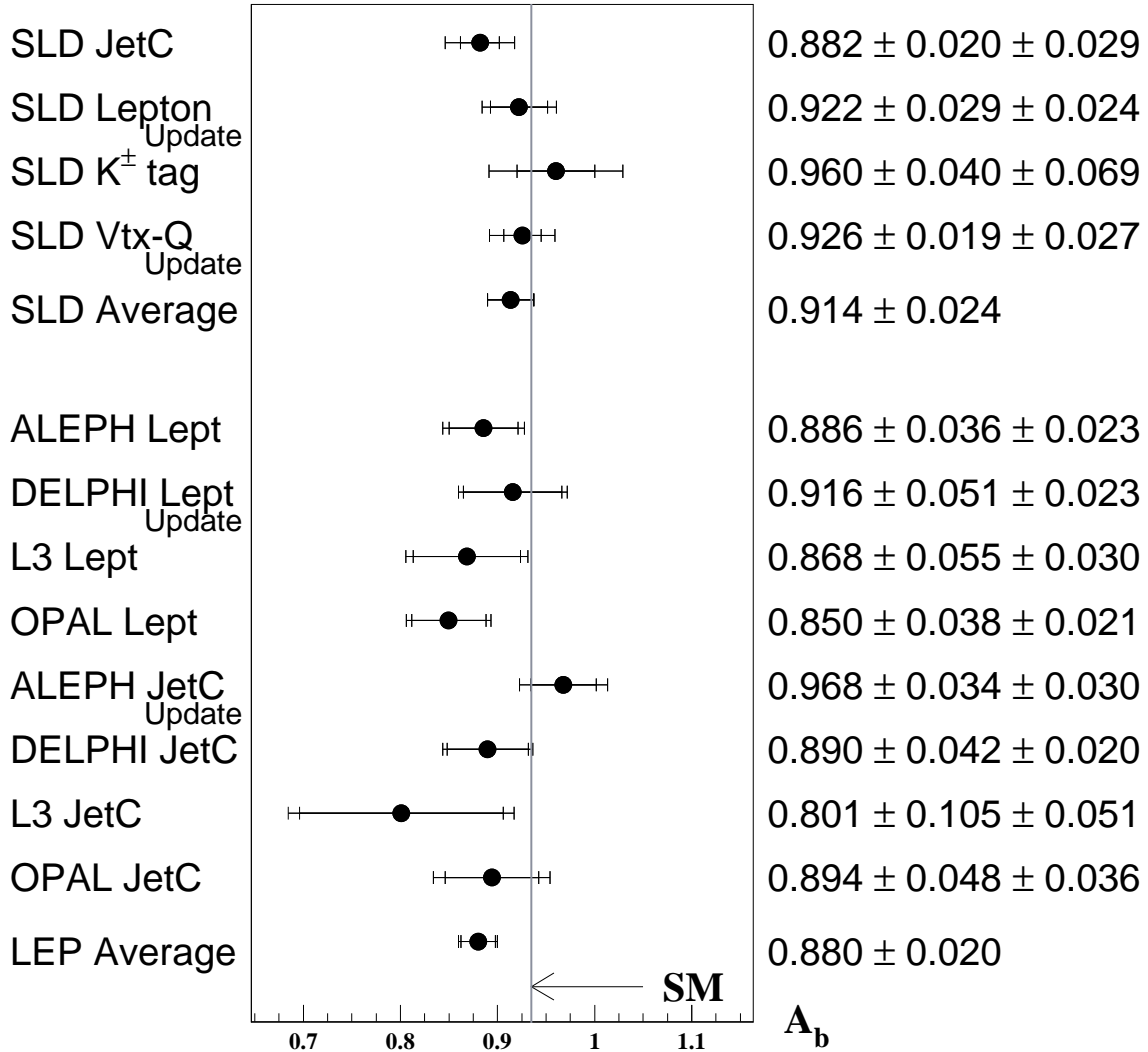}
\end{center}
\caption{The world $A_b$ measurements (Summer 2000).  LEP
measurements are derived from $A_b = 4A_{FB}^{0,b}$/$\left( 3A_e
\right)$ using $A_e = 0.1500\pm0.0016$ (the combined SLD $A_{LR}$
and LEP $A_{lepton}$).
}
\label{fig1}
\end{figure}

Fig.\ref{fig1} shows the preliminary results from the SLD and LEP
measurements, where the LEP
measurements are derived from $A_b = 4A_{FB}^{0,b}$/$\left( 3A_e
\right)$ using $A_e = 0.1500\pm0.0016$ (the combined SLD $A_{LR}$
and LEP $A_{lepton}$). 
The combined preliminary SLD result for $A_b$ is obtained as 
$A_b = 0.914\pm0.024 .$

\section{$A_c$ measurements}
At the SLD, four different techniques are used to measure the $A_c$:
1) inclusive charm-asymmetry measurement with kaon charge and vertex
charge, 
2) lepton, 
3) exclusively reconstructed $D^{\ast}$ and $D$-mesons, and
4) using $P_T$ spectrum of soft-pion from $D^*$.

In the inclusive charm analysis, $c$-quarks are tagged using 
intermediate $P_T$-corrected-mass vertices\cite{SLDAcinc}. 
It provides 82\% purity and 29\% efficiency 
for $Z^0\rightarrow c\bar{c}$ events. 
A $b$ veto is applied to reject any event with
high vertex mass in either hemisphere. For the
hemispheres with a secondary vertex, a secondary track identified as 
$K^\pm$ from the CRID, or a non-zero vertex charge, is used to sign the 
charm quark direction.
The background is mostly $b$ events and its fraction is constrained by 
the double-tag calibration.
This analysis has significantly high statistical power 
and the systematic errors are still very much under control. 

We also measure the charm asymmetry with traditional technique 
using electrons and muons which not only tag the $c$ events 
but also determine the $c$-quark direction from the lepton\cite{SLDAbAclept}.

The exclusive reconstruction of charmed mesons provide the cleanest
technique for the charm-asymmetry measurements\cite{SLDAcD*}.
We use four decay modes to identify $D^{\ast+}$:
the decay $D^{\ast+} \rightarrow \pi_s^+ D^0 $ followed by 
$D^0 \rightarrow K^- \pi^+$,       
$D^0 \rightarrow K^- \pi^+ \pi^0$  (Satellite resonance),
$D^0 \rightarrow K^- \pi^+ \pi^- \pi^+$, or
$D^0 \rightarrow K^- l^+ \nu_l$ ($l=$e or $\mu$).
We also identify $D^+$ and $D^0$ mesons via the decay of
$D^+ \rightarrow K^- \pi^+ \pi^+$ and        
$D^0 \rightarrow K^- \pi^+$ (not from $D^{\ast+}$).       
In this analysis, we reject $Z^0 \rightarrow b\bar{b}$
events using $P_T$-corrected mass of the reconstructed vertices.
The random-combinatoric background can be estimated 
from the mass sidebands. 

The soft-pions from the decay $D^{\ast+}\rightarrow D^0 \pi_s^+$
are also used to tag $c$-quarks\cite{SLDAcD*}.
To determine the $D^{\ast}$ direction, charged tracks and neutral
clusters are clustered into jets.
We also reject the $b\bar{b}$ background using $P_T$-corrected-mass
of reconstructed vertices.
Using the momenta transverse to the jet axis ($P_T$) for tracks, 
we select the soft-pion candidates which have small $P_T$ value.
The largest systematic uncertainty is the choice of the
background $P_T$ shape.

\begin{figure}
\begin{center}
\epsfysize 14cm
\epsfbox{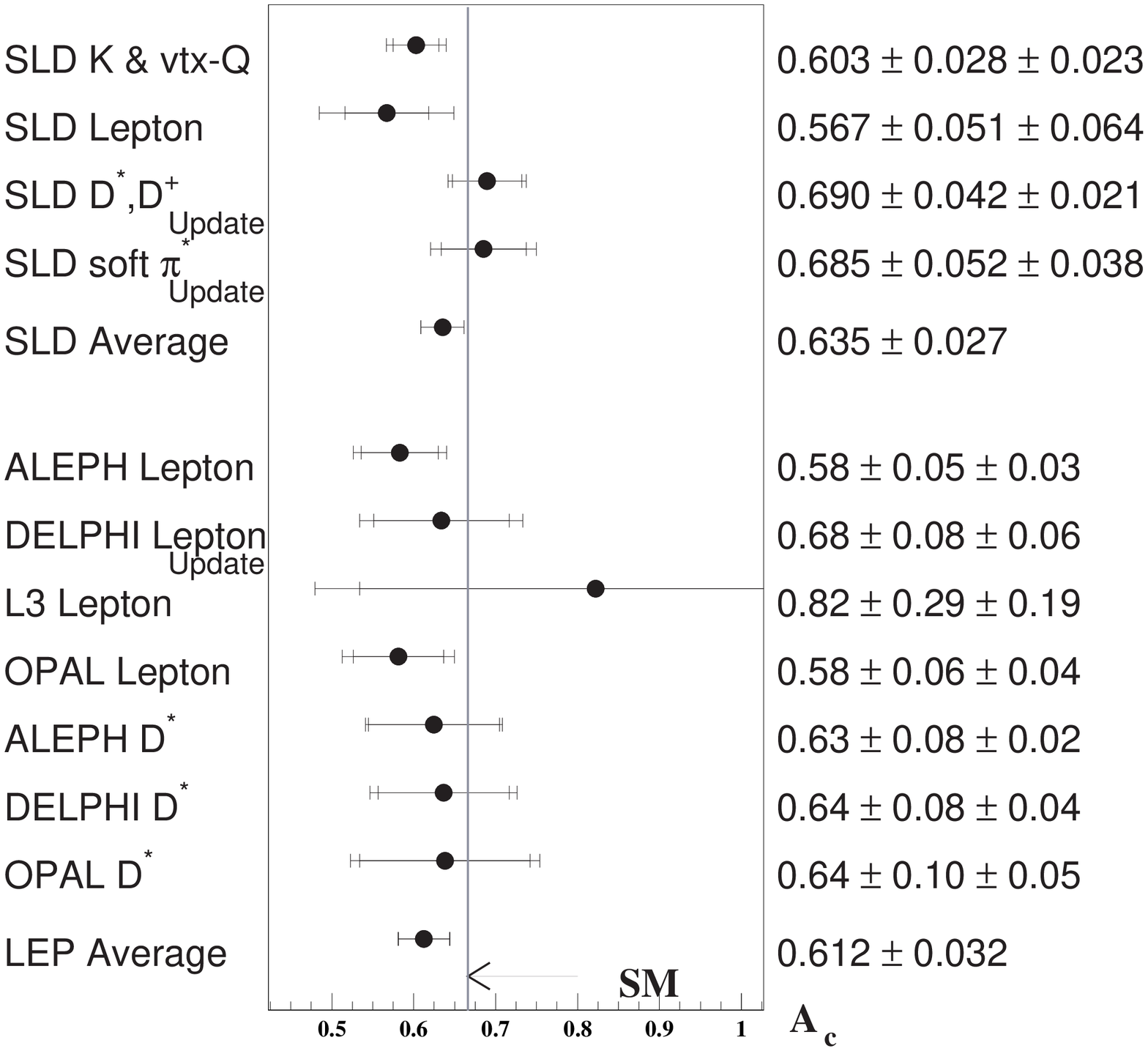}
\end{center}
\caption{The world $A_c$ measurements (Summer 2000).  LEP
measurements are derived from $A_c = 4A_{FB}^{0,c}$/$\left( 3A_e
\right)$ using $A_e = 0.1500\pm0.0016$ (the combined SLD $A_{LR}$
and LEP $A_{lepton}$).
}
\label{fig2}
\end{figure}

Fig.\ref{fig2} shows the preliminary results from the SLD and LEP
measurements.
The combined preliminary SLD result for $A_c$ is obtained as 
$A_c = 0.635\pm0.027 .$

\section{Measurement of $A_s$}
In this analysis, we use high-momentum strange particles\cite{SLDAs}.
We require both event hemispheres have $K^\pm$ with 
$p>9$ GeV/$c$, or $K^0_s$ with $p>5$ GeV/$c$. 
CRID is used to identify $K^\pm$.
To determine the $s$-quark charge, 
we require at least one hemisphere have $K^\pm$.
The heavy-quark background are rejected by identifying $B$ and $D$
decay vertices.
We obtain 66\% purity for 
$Z^0\rightarrow s\bar{s}$ events.

From the 1993-98 SLD data, we get the result of 
$A_s = 0.895\pm0.066(stat.)\pm0.062(sys.)$.
As a test of $d$-type quark universality, we compare
it with the SLD combined $A_b$ measurement:
$ A_b / A_s = 1.02\pm0.10. $
These are consistent within the error.

\section{Conclusion}
SLD produces world class measurements of parity-violation parameters.
The SLD measurements of $A_c$ and $A_s$ are now the most 
precise single measurements in the world. 
The measured $A_b$, $A_c$ and $A_s$ results are consistent with the 
Standard Model.


\section*{$^{**}$List of Authors} 
%
%
%
\begin{center}

\def\iAOMORI{$^{(1)}$}
\def\iBRI{$^{(2)}$}
\def\iBRUN{$^{(3)}$}
\def\iBU{$^{(4)}$}
\def\iCOLO{$^{(5)}$}
\def\iCSU{$^{(6)}$}
\def\iFERR{$^{(7)}$}
\def\iFRAS{$^{(8)}$}
\def\iJHU{$^{(9)}$}
\def\iLBL{$^{(10)}$}
\def\iMASS{$^{(11)}$}
\def\iMISSI{$^{(12)}$}
\def\iMIT{$^{(13)}$}
\def\iMOSCOW{$^{(14)}$}
\def\iNAGO{$^{(15)}$}
\def\iOREG{$^{(16)}$}
\def\iOXF{$^{(17)}$}
\def\iPERU{$^{(18)}$}
\def\iRAL{$^{(19)}$}
\def\iRUTG{$^{(20)}$}
\def\iSLAC{$^{(21)}$}
\def\iSOONG{$^{(22)}$}
\def\iTENN{$^{(23)}$}
\def\iTOHO{$^{(24)}$}
\def\iUCSB{$^{(25)}$}
\def\iUCSC{$^{(26)}$}
\def\iVAND{$^{(27)}$}
\def\iWASH{$^{(28)}$}
\def\iWISC{$^{(29)}$}
\def\iYALE{$^{(30)}$}

\baselineskip=.75\baselineskip 

\mbox{Kenji Abe\unskip,\iNAGO}
\mbox{Koya Abe\unskip,\iTOHO}
\mbox{T. Abe\unskip,\iSLAC}
\mbox{I. Adam\unskip,\iSLAC}
\mbox{H. Akimoto\unskip,\iSLAC}
\mbox{D. Aston\unskip,\iSLAC}
\mbox{K.G. Baird\unskip,\iMASS}
\mbox{C. Baltay\unskip,\iYALE}
\mbox{H.R. Band\unskip,\iWISC}
\mbox{T.L. Barklow\unskip,\iSLAC}
\mbox{J.M. Bauer\unskip,\iMISSI}
\mbox{G. Bellodi\unskip,\iOXF}
\mbox{R. Berger\unskip,\iSLAC}
\mbox{G. Blaylock\unskip,\iMASS}
\mbox{J.R. Bogart\unskip,\iSLAC}
\mbox{G.R. Bower\unskip,\iSLAC}
\mbox{J.E. Brau\unskip,\iOREG}
\mbox{M. Breidenbach\unskip,\iSLAC}
\mbox{W.M. Bugg\unskip,\iTENN}
\mbox{D. Burke\unskip,\iSLAC}
\mbox{T.H. Burnett\unskip,\iWASH}
\mbox{P.N. Burrows\unskip,\iOXF}
\mbox{A. Calcaterra\unskip,\iFRAS}
\mbox{R. Cassell\unskip,\iSLAC}
\mbox{A. Chou\unskip,\iSLAC}
\mbox{H.O. Cohn\unskip,\iTENN}
\mbox{J.A. Coller\unskip,\iBU}
\mbox{M.R. Convery\unskip,\iSLAC}
\mbox{V. Cook\unskip,\iWASH}
\mbox{R.F. Cowan\unskip,\iMIT}
\mbox{G. Crawford\unskip,\iSLAC}
\mbox{C.J.S. Damerell\unskip,\iRAL}
\mbox{M. Daoudi\unskip,\iSLAC}
\mbox{S. Dasu\unskip,\iWISC}
\mbox{N. de Groot\unskip,\iBRI}
\mbox{R. de Sangro\unskip,\iFRAS}
\mbox{D.N. Dong\unskip,\iMIT}
\mbox{M. Doser\unskip,\iSLAC}
\mbox{R. Dubois\unskip,\iSLAC}
\mbox{I. Erofeeva\unskip,\iMOSCOW}
\mbox{V. Eschenburg\unskip,\iMISSI}
\mbox{E. Etzion\unskip,\iWISC}
\mbox{S. Fahey\unskip,\iCOLO}
\mbox{D. Falciai\unskip,\iFRAS}
\mbox{J.P. Fernandez\unskip,\iUCSC}
\mbox{K. Flood\unskip,\iMASS}
\mbox{R. Frey\unskip,\iOREG}
\mbox{E.L. Hart\unskip,\iTENN}
\mbox{K. Hasuko\unskip,\iTOHO}
\mbox{S.S. Hertzbach\unskip,\iMASS}
\mbox{M.E. Huffer\unskip,\iSLAC}
\mbox{X. Huynh\unskip,\iSLAC}
\mbox{M. Iwasaki\unskip,\iOREG}
\mbox{D.J. Jackson\unskip,\iRAL}
\mbox{P. Jacques\unskip,\iRUTG}
\mbox{J.A. Jaros\unskip,\iSLAC}
\mbox{Z.Y. Jiang\unskip,\iSLAC}
\mbox{A.S. Johnson\unskip,\iSLAC}
\mbox{J.R. Johnson\unskip,\iWISC}
\mbox{R. Kajikawa\unskip,\iNAGO}
\mbox{M. Kalelkar\unskip,\iRUTG}
\mbox{H.J. Kang\unskip,\iRUTG}
\mbox{R.R. Kofler\unskip,\iMASS}
\mbox{R.S. Kroeger\unskip,\iMISSI}
\mbox{M. Langston\unskip,\iOREG}
\mbox{D.W.G. Leith\unskip,\iSLAC}
\mbox{V. Lia\unskip,\iMIT}
\mbox{C. Lin\unskip,\iMASS}
\mbox{G. Mancinelli\unskip,\iRUTG}
\mbox{S. Manly\unskip,\iYALE}
\mbox{G. Mantovani\unskip,\iPERU}
\mbox{T.W. Markiewicz\unskip,\iSLAC}
\mbox{T. Maruyama\unskip,\iSLAC}
\mbox{A.K. McKemey\unskip,\iBRUN}
\mbox{R. Messner\unskip,\iSLAC}
\mbox{K.C. Moffeit\unskip,\iSLAC}
\mbox{T.B. Moore\unskip,\iYALE}
\mbox{M. Morii\unskip,\iSLAC}
\mbox{D. Muller\unskip,\iSLAC}
\mbox{V. Murzin\unskip,\iMOSCOW}
\mbox{S. Narita\unskip,\iTOHO}
\mbox{U. Nauenberg\unskip,\iCOLO}
\mbox{H. Neal\unskip,\iYALE}
\mbox{G. Nesom\unskip,\iOXF}
\mbox{N. Oishi\unskip,\iNAGO}
\mbox{D. Onoprienko\unskip,\iTENN}
\mbox{L.S. Osborne\unskip,\iMIT}
\mbox{R.S. Panvini\unskip,\iVAND}
\mbox{C.H. Park\unskip,\iSOONG}
\mbox{I. Peruzzi\unskip,\iFRAS}
\mbox{M. Piccolo\unskip,\iFRAS}
\mbox{L. Piemontese\unskip,\iFERR}
\mbox{R.J. Plano\unskip,\iRUTG}
\mbox{R. Prepost\unskip,\iWISC}
\mbox{C.Y. Prescott\unskip,\iSLAC}
\mbox{B.N. Ratcliff\unskip,\iSLAC}
\mbox{J. Reidy\unskip,\iMISSI}
\mbox{P.L. Reinertsen\unskip,\iUCSC}
\mbox{L.S. Rochester\unskip,\iSLAC}
\mbox{P.C. Rowson\unskip,\iSLAC}
\mbox{J.J. Russell\unskip,\iSLAC}
\mbox{O.H. Saxton\unskip,\iSLAC}
\mbox{T. Schalk\unskip,\iUCSC}
\mbox{B.A. Schumm\unskip,\iUCSC}
\mbox{J. Schwiening\unskip,\iSLAC}
\mbox{V.V. Serbo\unskip,\iSLAC}
\mbox{G. Shapiro\unskip,\iLBL}
\mbox{N.B. Sinev\unskip,\iOREG}
\mbox{J.A. Snyder\unskip,\iYALE}
\mbox{H. Staengle\unskip,\iCSU}
\mbox{A. Stahl\unskip,\iSLAC}
\mbox{P. Stamer\unskip,\iRUTG}
\mbox{H. Steiner\unskip,\iLBL}
\mbox{D. Su\unskip,\iSLAC}
\mbox{F. Suekane\unskip,\iTOHO}
\mbox{A. Sugiyama\unskip,\iNAGO}
\mbox{S. Suzuki\unskip,\iNAGO}
\mbox{M. Swartz\unskip,\iJHU}
\mbox{F.E. Taylor\unskip,\iMIT}
\mbox{J. Thom\unskip,\iSLAC}
\mbox{E. Torrence\unskip,\iMIT}
\mbox{T. Usher\unskip,\iSLAC}
\mbox{J. Va'vra\unskip,\iSLAC}
\mbox{R. Verdier\unskip,\iMIT}
\mbox{D.L. Wagner\unskip,\iCOLO}
\mbox{A.P. Waite\unskip,\iSLAC}
\mbox{S. Walston\unskip,\iOREG}
\mbox{A.W. Weidemann\unskip,\iTENN}
\mbox{E.R. Weiss\unskip,\iWASH}
\mbox{J.S. Whitaker\unskip,\iBU}
\mbox{S.H. Williams\unskip,\iSLAC}
\mbox{S. Willocq\unskip,\iMASS}
\mbox{R.J. Wilson\unskip,\iCSU}
\mbox{W.J. Wisniewski\unskip,\iSLAC}
\mbox{J.L. Wittlin\unskip,\iMASS}
\mbox{M. Woods\unskip,\iSLAC}
\mbox{T.R. Wright\unskip,\iWISC}
\mbox{R.K. Yamamoto\unskip,\iMIT}
\mbox{J. Yashima\unskip,\iTOHO}
\mbox{S.J. Yellin\unskip,\iUCSB}
\mbox{C.C. Young\unskip,\iSLAC}
\mbox{H. Yuta\unskip.\iAOMORI}

\it
\vskip \baselineskip                   
\vskip \baselineskip
\baselineskip=.75\baselineskip   
\iAOMORI
  Aomori University, Aomori , 030 Japan, \break
\iBRI
  University of Bristol, Bristol, United Kingdom, \break
\iBRUN
  Brunel University, Uxbridge, Middlesex, UB8 3PH United Kingdom, \break
\iBU
  Boston University, Boston, Massachusetts 02215, \break
\iCOLO
  University of Colorado, Boulder, Colorado 80309, \break
\iCSU
  Colorado State University, Ft. Collins, Colorado 80523, \break
\iFERR
  INFN Sezione di Ferrara and Universita di Ferrara, I-44100 Ferrara, Italy, \break
\iFRAS
  INFN Lab. Nazionali di Frascati, I-00044 Frascati, Italy, \break
\iJHU
  Johns Hopkins University,  Baltimore, Maryland 21218-2686, \break
\iLBL
  Lawrence Berkeley Laboratory, University of California, Berkeley, California 94720, \break
\iMASS
  University of Massachusetts, Amherst, Massachusetts 01003, \break
\iMISSI
  University of Mississippi, University, Mississippi 38677, \break
\iMIT
  Massachusetts Institute of Technology, Cambridge, Massachusetts 02139, \break
\iMOSCOW
  Institute of Nuclear Physics, Moscow State University, 119899, Moscow Russia, \break
\iNAGO
  Nagoya University, Chikusa-ku, Nagoya, 464 Japan, \break
\iOREG
  University of Oregon, Eugene, Oregon 97403, \break
\iOXF
  Oxford University, Oxford, OX1 3RH, United Kingdom, \break
\iPERU
  INFN Sezione di Perugia and Universita di Perugia, I-06100 Perugia, Italy, \break
\iRAL
  Rutherford Appleton Laboratory, Chilton, Didcot, Oxon OX11 0QX United Kingdom, \break
\iRUTG
  Rutgers University, Piscataway, New Jersey 08855, \break
\iSLAC
  Stanford Linear Accelerator Center, Stanford University, Stanford, California 94309, \break
\iSOONG
  Soongsil University, Seoul, Korea 156-743, \break
\iTENN
  University of Tennessee, Knoxville, Tennessee 37996, \break
\iTOHO
  Tohoku University, Sendai 980, Japan, \break
\iUCSB
  University of California at Santa Barbara, Santa Barbara, California 93106, \break
\iUCSC
  University of California at Santa Cruz, Santa Cruz, California 95064, \break
\iVAND
  Vanderbilt University, Nashville,Tennessee 37235, \break
\iWASH
  University of Washington, Seattle, Washington 98105, \break
\iWISC
  University of Wisconsin, Madison,Wisconsin 53706, \break
\iYALE
  Yale University, New Haven, Connecticut 06511. \break

\rm
%

\end{center}

%
%


\end{document}